# Mesotuned WIMPs and avoided deconfinement

Michael Nee
*Rudolf Peierls Centre for Theoretical Physics, University of Oxford,
Parks Road, Oxford OX1 3PU, United Kingdom*

We discuss the production of dark matter with mass of order 10–100 TeV and $\mathcal{O}(1)$ couplings to the standard model, a scenario we refer to as the "mesotuned WIMP." Given the lack of new physics observed at the LHC, indicating we live in a world with some degree of fine-tuning, this scale is a natural scale to expect new particles to appear. However, dark matter with mass in this range and large couplings is difficult to produce with the observed relic abundance via the usual mechanisms. Here we discuss the nonthermal production of dark matter and show that this mechanism can produce the observed dark matter in this mass range, but requires a specific relationship between the dark matter mass and the reheat temperature. It is shown that the avoided deconfinement model removes the dependence on the reheat temperature and predicts a dark matter candidate that can be tested at the Cherenkov Telescope Array and direct detection experiments, offering a natural setting for mesotuned WIMP dark matter.

## I. INTRODUCTION

The most widely studied paradigm for dark matter (DM) production is WIMP freeze-out. This is motivated by the so-called WIMP miracle, the discovery that for weak-scale masses and $\mathcal{O}(1)$ couplings to the SM the freeze-out mechanism produces the correct relic abundance of dark matter, thus pointing to a scale where new particles are already expected to exist from naturalness arguments (see, e.g., [1]). However, null results from the LHC may point to the fact that we live in a "mesotuned" universe with some degree of fine-tuning. In this case a mechanism that renders the Higgs mass natural may not appear until the scale of 10–100 TeV or higher [2,3]. If we take this indication at face value, then we might expect dark matter to have a mass set at the same scale and $\mathcal{O}(1)$ couplings to the Standard Model (SM), a DM candidate we term the "mesotuned WIMP."

This scenario is testable in current and upcoming experiments. In this mass range direct detection searches are limited by the lower number density of DM particles, but experiments such as PandaX2 [4,5] and Xenon1T [6,7] still place strong limits on the nucleon scattering cross section. Cherenkov telescopes such as CTA [8–11], H.E.S.S. [12], and HAWC [13,14] are also sensitive to the gamma ray emission from DM annihilations for masses up to $m_\chi \sim \mathcal{O}(10^2)$ TeV. This further motivates the study of production mechanisms that can lead to a viable DM candidate with masses at this scale and large SM couplings.

However, DM with these properties is difficult to produce with the correct relic abundance. Unitarity bounds limit the maximum mass for particles produced via freeze-out to be around $\sim 100$ TeV [15,16]. This can be circumvented if there is significant entropy production in the early universe [17,18] or the dark sector contains extra degrees of freedom with approximately degenerate masses [19–24]. An alternative production mechanism for particles with masses in this range is the freeze-in mechanism, but this typically requires very small couplings

$$\lambda \sim 10^{-10} \quad (1)$$

in order to prevent the DM from thermalizing [25–30]. This makes freeze-in models generally difficult to observe, although they may have interesting signatures at colliders [31,32].

An alternative to these paradigms that allows generation of the correct DM relic abundance is nonthermal production [33,34]. This mechanism bears some similarity to the freeze-in mechanism as the DM never reaches thermal equilibrium, but can be achieved with $\mathcal{O}(1)$ couplings if the DM is never relativistic. For a $2 \leftrightarrow 2$ process, the requirement on the coupling then becomes

$$\lambda e^{-m_\chi/T_r} \sim 10^{-10} \quad (2)$$

for DM mass $m_\chi$ and maximum temperature of the universe $T_r$. This mechanism generates a relic abundance that is independent of the DM mass for a fixed ratio $m_\chi/T_r$ and has been considered as a production mechanism for superheavy dark matter [33–41]. While this mechanism can explain the correct relic abundance, for DM with $m_\chi \sim$ 10–100 TeV it requires a somewhat *ad hoc* assumption about the maximum temperature obtained in the universe:

$$T_r \simeq \frac{m_\chi}{23 + \log(\lambda)}. \tag{3}$$

Depending on the precise value of $m_\chi$ this may make baryogenesis difficult to accommodate in these models if $T_r$ is below the electroweak scale.

A model that circumvents this issue is the avoided deconfinement (AD) model [42]. The AD model is a modification to the Randall-Sundrum (RS) [43] model stabilized by the Goldberger-Wise mechanism [44]. The high-temperature phase of the RS model is described by the AdS-Schwarzschild solution, and the phase transition is heavily suppressed due to the approximate conformal symmetry of the model [45]. The AD model offers a resolution to this cosmological problem by rendering the RS phase metastable up to very high temperatures. The maximum temperature up to which the AD model remains metastable depends on the model parameters; for the benchmark points presented in Ref. [42] the maximum possible temperature is $T_r \sim 5 \times 10^{11}$ GeV.

The RS model also offers an elegant solution to the hierarchy problem with the weak scale set by the radion vacuum expectation value (VEV), $\mu$. Collider bounds from the searches for Kaluza Klein (KK) modes set this scale to be $\mu \gtrsim 4.25$ TeV [46,47], meaning some amount of fine-tuning is required in the model as a solution to the hierarchy problem. Thus, the RS model naturally predicts particles in the 10–100 TeV range which could serve as dark matter candidates.

Additionally, the AD model introduces a temperature dependence of the radion expectation value at high $T$,

$$\frac{\mu(T)}{\mu(0)} \propto \left(\frac{T}{T_c}\right)^{1-\epsilon}, \tag{4}$$

for $\epsilon \ll 1$ and $T_c$ the critical temperature of the model that marks the onset of the high-temperature regime, which necessarily satisfies $T_c < \mu(0)$. As the radion sets the scale of physics on the IR brane, all IR scales exhibit the same temperature dependence, so that at high temperature

$$\frac{m_\chi(T)}{T} \propto \left(\frac{T_c}{T}\right)^\epsilon \tag{5}$$

for IR-localized fields. Therefore, if $m_\chi(T_c)/T_c > 1$, a reasonable expectation in the RS model as $T_c < \mu(0)$, $\chi$ remains nonrelativistic up to very high temperatures. This makes the model a natural setting for DM with masses $m_\chi \sim$ 10–100 TeV produced nonthermally, while allowing for large reheat temperatures.

In this work we consider the nonthermal production of mesotuned WIMPs—DM candidates with $\mathcal{O}(1)$ SM couplings and masses in the 1–100 TeV range. In Sec. II we present the nonthermal production mechanism assuming a low reheat temperature and radiation dominated universe. We then show how this is modified in the AD model in Sec. III and go through a specific example of scalar dark matter produced via the radion portal in Sec. III A. Experimental signatures of the model are discussed in Sec. IV.

## II. NONTHERMAL DARK MATTER PRODUCTION

The usual freeze-in mechanism [30] applies to particles $\chi$ that have a negligible initial abundance and a highly suppressed coupling $\lambda$ to particles in the thermal bath. This suppressed interaction produces an abundance of $\chi$ particles that remains below the equilibrium value and freezes-in at a temperature such that $x = m_\chi/T \sim 2$–5. Below this temperature the production and annihilation rates for $\chi$ particles are exponentially suppressed and the comoving number density remains constant.

In contrast to the freeze-out scenario, the $\chi$ abundance produced by freeze-in increases with $\lambda$, and typically values of $\lambda \sim 10^{-10}$ are required to produce the observed DM abundance. A coupling of this order also means that these models are difficult to detect. However, if the DM candidate $\chi$ has a mass ($m_\chi$) larger than the reheat temperature ($T_r$), then the freeze-in mechanism can produce the correct relic abundance for $\mathcal{O}(1)$ couplings between $\chi$ and the thermal plasma. This works as the $\chi$ production process is suppressed by the Boltzmann factor $e^{-m_\chi/T}$, so the effective coupling that governs the freeze-in process is not $\lambda$ but

$$\lambda e^{-x_r}, \tag{6}$$

where $x_r = m_\chi/T_r \gtrsim 1$. The requirement for the correct relic abundance then becomes $\lambda e^{-x_r} \sim 10^{-10}$, which can be easily achieved for $\lambda \sim 1$ and $x_r \sim 25$.

### A. Boltzmann equations

To illustrate this we consider a model where the dominant process contributing to the $\chi$ abundance is

$$\chi\chi \leftrightarrow \phi\phi, \tag{7}$$

for a thermalized species $\phi$ which is taken to be massless. We will parametrize the matrix element for this process as $\mathcal{M} = \lambda$ for some effective coupling $\lambda$. If the process (7) is governed by a renormalizable operator, through, e.g., a Higgs portal, then $\lambda$ is simply the coupling in the Lagrangian. If $\chi$, $\phi$, or both are fermions, the process comes from a higher-dimensional operator, as in models with a heavy mediator [48,49]. In that case, the effective coupling is related to the tree-level coupling by

$$\lambda = \lambda_{\text{tree}} \left(\frac{m_\chi}{M_{\text{uv}}}\right)^{d-4}, \quad (8)$$

where $d$ is the dimension of the operator and $M_{\text{uv}} > m_\chi$ a UV scale which is integrated out.

The number density of $\chi$ particles in the early universe is governed by the Boltzmann equation:

$$\dot{n}_\chi + 3Hn_\chi = \langle\sigma v\rangle(n_{\text{eq}}^2 - n_\chi^2), \quad (9)$$

where $n_{\text{eq}}$ is the equilibrium number density. The thermally averaged cross section is given by the integral [50,51]

$$n_{\text{eq}}^2 \langle\sigma v\rangle = \frac{g_\chi^2 T}{512\pi^5} \int_{4m_\chi^2}^\infty ds |\mathcal{M}|^2 K_1\left(\frac{\sqrt{s}}{T}\right)\sqrt{s - 4m_\chi^2}, \quad (10)$$

where $K_1$ is a modified Bessel function and $g_\chi$ is the number of degrees of freedom of the field $\chi$. Extracting the parametric dependence for large $x$, we find (refer to Appendix A for details)

$$n_{\text{eq}}^2 \langle\sigma v\rangle \simeq \frac{g_\chi^2 T^4}{256\pi^4} \lambda^2 x e^{-2x}. \quad (11)$$

Using this result and rewriting Eq. (9) in terms of the yield $Y_\chi = n_\chi/s$, we find

$$\frac{dY_\chi}{dx} = \frac{45 M_P \lambda^2 x e^{-2x} g_\chi^2}{1.66 \times 512\pi^6 g_{*S} \sqrt{g_*} m_\chi}\left(1 - \frac{Y_\chi^2}{Y_{\text{eq}}^2}\right). \quad (12)$$

We can then integrate from $x = x_r$ to $x \to \infty$ to get

$$Y_\chi(T \to 0) = \frac{45 \lambda^2 g_\chi^2 M_P}{1.66 \times 2048\pi^6 g_{*S} \sqrt{g_*} m_\chi}(1 + 2x_r)e^{-2x_r}, \quad (13)$$

leading to a present-day abundance of

$$\Omega_\chi h^2 = 0.413 \times \left(\frac{\lambda e^{-x_r}}{10^{-10}}\right)^2 (1 + 2x_r), \quad (14)$$

assuming $g_\chi = 2$.

Here we note that this depends on the mass only through the parameter $x_r = m_\chi/T_r$. This mechanism therefore offers a mechanism for DM production over a wide mass range, although it requires a specific ratio of the mass to reheat temperature. For DM in the mesotuned range with $\mathcal{O}(1)$ couplings this points to a reheat temperature in the range 40 GeV to 4 TeV. A reheat temperature in this range makes a baryogenesis mechanism difficult to accommodate, although production of a baryon asymmetry in models with a low reheat temperature is possible [52].

Here we also note the possibility of a period of matter domination just after inflation, as may be the case in models with a reheat temperature at the TeV scale or below [34,36]. In this case the decay of a scalar field reheats the universe after inflation. If the decay is rapid, then the reheat temperature is given in terms of the inflationary hubble scale $H_I$ as

$$T_r = \sqrt{M_P H_I}. \quad (15)$$

In this case the above analysis holds although this requires a very low Hubble scale during inflation.

If instead the scalar field decays slowly, then $H_I$ can be larger than the Hubble rate at $T_r$. The temperature in this case increases to a maximum of $T_{\text{max}}$ before cooling with the scale factor $a$ as $T \propto a^{-3/8}$ until reaching $T_r$ at the start of the radiation dominated era. For large $x_{\text{max}} = m_\chi/T_{\text{max}}$ the DM abundance in this case is given by replacing [38,41]

$$e^{-2x_r} \to \left(\frac{T_r}{T_{\text{max}}}\right)^8 e^{-2x_{\text{max}}}, \quad (16)$$

in Eq. (14), up to an $\mathcal{O}(1)$ prefactor. This leads to a suppressed present day density compared to the case considered in this work. There is also the possibility that the $\chi$ abundance is produced through the decay of the inflaton [33–37,39] or moduli [40]. In both cases an additional dependence on the details of inflation is introduced through the parameter $T_{\text{max}}$, and as such we do not consider these cases here, instead assuming instantaneous reheating after inflation.

As discussed in the following section, the AD model provides a scenario where DM is produced in the same way without having to assume a low reheat temperature, a scenario where the approximation of instantaneous decay is more consistent with models of inflation.

## III. AVOIDED DECONFINEMENT

In this section, a brief overview of the AD mechanism is presented, with emphasis on the features that lead to new possibilities for DM production. For the full details of the model, the reader is referred to Ref. [42]. The relevant feature of the mechanism for our work is that all scales on the IR brane are set by the radion VEV, denoted $\mu(T)$, which increases approximately linearly with $T$ above a critical temperature $T_c$ which marks the onset of the AD phase.

The AD model requires that the universe exit inflation in the RS phase then reheats to a maximum temperature $T_r$. At high temperatures the RS phase is rendered metastable by a symmetry nonrestoration mechanism that leads to the radion VEV scaling with temperature as

$$\mu(T) = \mu(0)\left(\frac{T}{T_c}\right)^{\frac{1}{1+\epsilon}}. \quad (17)$$

$T_c$ is the temperature that marks the onset of the AD phase[1] and $\epsilon$ is a small parameter related to the explicit breaking of conformal symmetry. Below $T_c$ the radion VEV settles to its zero-temperature value and the cosmology is that of the usual RS model. $\mu(T=0)$ sets the weak scale and is taken to be of order 10–100 TeV, alleviating the hierarchy problem and satisfying all bounds from colliders [46,47,53,54], while the value of $T_c$ depends on the parameters of the model but must satisfy $T_c < \mu(0)$.

The temperature dependence of the radion VEV leads to all dimensionful parameters on the IR brane scaling with temperature in a similar way. In particular, the mass of an IR degree of freedom will have a temperature dependence of

$$m_{\text{ir}}(T) = m_{\text{ir}}(0)\frac{\mu(T)}{\mu(0)} = m_{\text{ir}}(0)\left(\frac{T}{T_c}\right)^{\frac{1}{1+\epsilon}}. \tag{18}$$

This implies that the ratio $x(T) = m_{\text{ir}}(T)/T$ changes by a factor of

$$\frac{x(T_r)}{x(T_c)} = \left(\frac{T_c}{T_r}\right)^{\frac{\epsilon}{1+\epsilon}}, \tag{19}$$

during the period of avoided deconfinement. Taking the benchmark parameters of [42], $T_r \sim 10^{11}$ GeV, $T_c = 16.6$ TeV, $\epsilon = 4.13 \times 10^{-2}$ gives

$$\frac{x(T_r)}{x(T_c)} = 0.53. \tag{20}$$

This allows for the nonthermal production of dark matter with a mass around the TeV scale while requiring only mild assumptions about the reheat temperature.[2] Any particle with mass set by the radion VEV which satisfies $m \gtrsim 2T_c$ will remain nonrelativistic throughout the entire thermal history due to the temperature dependence of the radion.

In comparison to typical nonthermal dark matter production, the yield in the AD model is enhanced due to the longer time the ratio $x$ remains constant. This is captured by the relation

$$\frac{dx}{dT} = -\frac{\epsilon x}{T}, \tag{21}$$

---

[1]The definition of $T_c$ in this work differs by the factor $c$ from the definition in [42].

[2]The maximum allowed reheat temperature in the AD model is bounded by the requirement that the RS phase remains metastable; for the benchmark points presented here this requires $T_R \lesssim 5 \times 10^{11}$ GeV.

as opposed to $dx/dT = -x/T$ in the usual case. This leads to a $1/\epsilon$ enhancement of the final abundance

$$\Omega_\chi h^2 = 0.1 \times \left(\frac{4.13 \times 10^{-2}}{\epsilon}\right)\left(\frac{\lambda e^{-x_r}}{10^{-11}}\right)^2 (1 + 2x_r) \tag{22}$$

giving the correct present-day abundance for slightly larger values of $x_r$ given fixed coupling $\lambda$.

### A. Radion portal dark matter

The canonically normalized radion field, $\varphi$, has couplings to IR degrees of freedom set by the broken conformal symmetry. The radion mass in the AD model is given by

$$m_\varphi^2 \simeq 0.1 \times \epsilon^{3/4}\mu \tag{23}$$

so it is parametrically below the scale set by $\mu$. Collider bounds on KK resonances imply that $\mu \gtrsim 4.25$ TeV, leading to a radion mass $m_\varphi \gtrsim 50$ GeV [46,47]. Any DM candidate in the AD model necessarily couples through the radion portal, so a minimal model is presented here where the radion couplings set the DM abundance [55–57]. The radion couplings to the SM can be determined by replacing mass scales $M$ in the SM Lagrangian with

$$M \to M\left(1 + \frac{\varphi}{F_\varphi}\right). \tag{24}$$

Neglecting a possible coupling of the Higgs to the Ricci scalar, there is also a coupling of the radion to the Higgs kinetic term

$$\frac{\varphi}{F_\varphi}(D_\mu H^\dagger)(D_\mu H). \tag{25}$$

Here $F_\varphi$ is the radion decay constant,

$$F_\varphi = \frac{N}{2\sqrt{2}\pi}\mu, \tag{26}$$

where $N$ is the large-$N$ parameter of the RS model. Below the electroweak scale and in unitary gauge, the interaction Lagrangian governing the radion interactions with the SM is

$$\mathcal{L}_{\varphi,\text{int}} = \frac{\varphi}{2F_\varphi}(\partial h)^2 - \sum_\psi \left(\frac{\varphi}{F_\varphi}\right)m_\psi\bar{\psi}\psi - \frac{5m_\varphi^2}{3F_\varphi}\varphi^3 - \frac{11}{24}\frac{m_\varphi^2}{F_\varphi^2}\varphi^4$$
$$+ \left(\frac{2\varphi}{F_\varphi} + \frac{\varphi^2}{F_\varphi^2}\right)\left(m_W^2 W_\mu^+ W^{-\mu} + \frac{m_Z^2}{2}Z_\mu Z^\mu - \frac{m_h^2}{2}h^2\right). \tag{27}$$

For the case of scalar $\chi$, the couplings to the radion are described by the Lagrangian

$$\mathcal{L}_{\varphi,\chi} = -\frac{m_\chi^2}{2}\left(1 + \frac{\varphi}{F_\varphi}\right)^2 \chi^2. \tag{28}$$

In the limit where $\chi$ is heavier than the SM fields and the radion, the dominant processes that contribute to freeze-out are the $\chi$ annihilations to the Higgs, radion, and longitudinal modes of the electroweak gauge bosons. In this limit the squared matrix elements for these processes are all equal and given by

$$|\mathcal{M}|^2 = 4\left(\frac{m_\chi}{F_\varphi}\right)^4. \tag{29}$$

The DM abundance is therefore governed by the cross section

$$\sigma v = \frac{\lambda_{\text{eff}}^2}{32\pi m_\chi^2}, \tag{30}$$

where we have defined the effective coupling:

$$\lambda_{\text{eff}} = 2\sqrt{5}\left(\frac{m_\chi}{F_\varphi}\right)^2, \tag{31}$$

which can be substituted for $\lambda$ in Eq. (22) to determine the relic abundance.

The avoided deconfinement model solves two issues with the nonthermal production process for DM in the mesotuned range. The RS model solves the hierarchy problem as the radion sets the scale of physics on the IR brane to be of order 10–100 TeV, predicting the existence of new particles with masses $m \sim c\mu(0)$ for $c \sim \mathcal{O}(1)$. Given $T_c < \mu(0)$, any new particle with $c \gtrsim 2$ will remain nonrelativistic for the thermal history of the universe, and stable particles can have a relic abundance set by a similar process to the one described here without having to assume a relationship between $T_r$ and $m_\chi$. Furthermore, the assumption of instantaneous reheating is more natural in the AD model as the reheat temperature can be large without $\chi$ becoming relativistic, meaning no assumptions about the Hubble scale of inflation are necessary.

## IV. EXPERIMENTAL SIGNATURES

In this section, the possible experimental signatures of mesotuned WIMPs are discussed. The experiments sensitive to the model depend on the final state of $\chi$ annihilations that set the relic abundance. If $\chi$ couples to nuclei, the strongest bounds come from direct detection, while if the production process comes from $\chi$ annihilations to $\tau$ or $W$ final states, then CTA and H.E.S.S. are sensitive to the gamma ray signal form $\chi$ annihilations in the galactic center.

Experiments sensitive to the $\gamma$ ray emission from the galactic center are most sensitive to $\chi$ annihilations to $\tau$ or $W$ final states. In the mass range $m_\chi \sim 10–100$ TeV, the projected bounds from CTA are of order [8–11]

$$\sigma v \lesssim 10^{-26}\text{–}10^{-25}\ \text{cm}^3\ \text{s}^{-1}. \tag{32}$$

H.E.S.S. has a similar sensitivity in the range $m_\chi \sim 1$–10 TeV, but the sensitivity falls off more rapidly at higher masses [12]. HAWC is also sensitive to gamma-ray emission but the bounds are above the cross sections predicted by the model considered in this work [13,14]. Indirect detection experiments are also sensitive to $\chi$ decays to a $b\bar{b}$ final state, but in this case direct detection offers the strongest constraints on the model. Compared to the $\chi\chi \to b\bar{b}$ cross section, the cross section for $\chi$ scattering with nuclei is further suppressed by a form factor of $f_b^2 \sim 10^{-4}$; however, PandaX2 [4,5] and Xenon1T [6,7] place strong bounds on the cross section to nuclei:

$$\sigma_{\chi N} \lesssim 3 \times 10^{-36}\ \text{cm}^3\ \text{s}^{-1}\left(\frac{m_\chi}{\text{TeV}}\right), \tag{33}$$

offering stronger constraints than indirect detection experiments.

### A. Constraints on radion portal DM

For the case of the scalar model with couplings via the radion portal presented in Sec. III A the dominant bounds come from CTA and direct detection experiments. The relevant cross sections are the annihilation to $W$ bosons

$$\sigma v(\chi\chi \to WW) = \frac{\lambda_{\text{eff}}^2}{80\pi m_\chi^2} \tag{34}$$

and scattering off nuclei [56]

$$\sigma(\chi N \to \chi N) \simeq \frac{F_N^2 \lambda_{\text{eff}}^2}{20\pi}\frac{m_N^2}{m_\chi^2 m_\varphi^4},$$
$$F_N \sim 0.6\ \text{GeV}, \tag{35}$$

which is derived from the interaction terms coupling the radion to quarks:

$$\mathcal{L} \supset -\sum_q \frac{m_q}{F_\varphi}\varphi\bar{q}q. \tag{36}$$

The bounds on the model are shown in Figs. 1 and 2 for fixed effective couplings $\lambda_{\text{eff}} = 1$ and $\lambda_{\text{eff}} = 0.1$, respectively. Direct detection constraints (blue shaded region) rule out the region of parameter space where both the radion and $\chi$ are light. At larger masses direct detection experiments lose sensitivity due to the $m_\chi^{-2} m_\varphi^{-4}$ dependence of the cross section. Lines of constant $N$ are shown in gray, defined implicitly as a function of $m_\chi$ from the requirement of constant $\lambda_{\text{eff}}$. Similarly, the dotted horizontal lines show the values of the radion VEV $\mu$ given the radion mass, and the vertical dashed lines show the critical temperature $T_c$ of the AD model required to generate the correct relic abundance. In both figures the reheat temperature $T_r$ is fixed to be $10^{10}$ GeV and $\epsilon$ set to $4.13 \times 10^{-2}$. For small $\epsilon$ the model is largely insensitive to $T_r$—increasing $T_r$ to the Planck scale results only in a shift of the value of $T_c$ required to produce the observed DM abundance by a factor of 2.

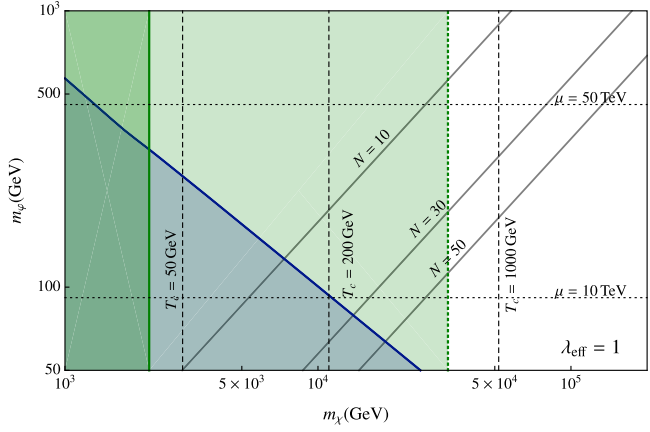

FIG. 1. Bounds on the parameter space for fixed $\lambda_{\text{eff}} = 1$. The blue shaded region is excluded by direct detection null results. The light green region shows the bounds from CTA assuming the maximum possible Sommerfeld enhancement, $S \sim 200$, while the dark green region shows the CTA bounds when Sommerfeld enhancement is negligible. The ratio $x_r$ required to generate the correct relic abundance is $x_r = 27$.

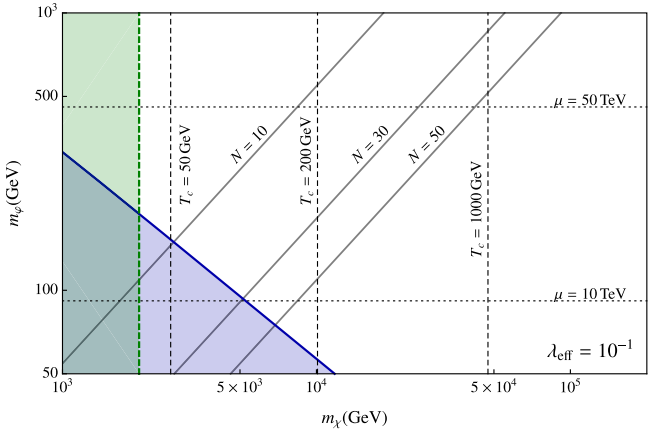

FIG. 2. Bounds on the parameter space for fixed $\lambda_{\text{eff}} = 0.1$. The blue shaded region is excluded by direct detection null results. The light green region shows the bounds from CTA assuming the maximum possible Sommerfeld enhancement, $S \sim 86$. The ratio $x_r$ required to generate the correct relic abundance is $x_r = 25$.

For $\lambda_{\text{eff}} \sim \mathcal{O}(1)$ CTA is sensitive to $\chi$ masses up to the multi-TeV range (green shaded regions), assuming the maximum sensitivity $(\sigma v)_{\chi W} < 10^{-26}$ cm$^3$ s$^{-1}$ across the mass range. The dark green region in Fig. 1 shows the CTA reach without considering possible Sommerfeld enhancement of the cross section [58], while the light green regions in Figs. 1 and 2 show the CTA sensitivity assuming the maximum possible Sommerfeld enhancement.[3] For $\lambda_{\text{eff}} \lesssim 10^{-1}$ CTA begins to lose sensitivity to the model even in the presence of a significant enhancement of the cross section.

The enhancement factor is derived by resumming ladder diagrams involving dilaton exchange between $\chi$ legs and can be approximated as [59–61]

$$S \simeq \frac{\pi}{\epsilon_v} \frac{\sinh\left(\frac{12\epsilon_v}{\pi\epsilon_\phi}\right)}{\cosh\left(\frac{12\epsilon_v}{\pi\epsilon_\phi}\right) - \cos\left(2\pi\sqrt{\frac{6}{\pi^2\epsilon_\phi} - \left(\frac{12\epsilon_v}{\pi\epsilon_\phi}\right)^2}\right)}, \quad (37)$$

where $\epsilon_v = \frac{2\pi v F_\varphi^2}{m_\chi^2}$, $\epsilon_\phi = \frac{2\pi m_\varphi F_\varphi^2}{m_\chi^3}$, and $v \sim 10^{-3}$ is the DM velocity. This factor is only significant at resonances in tuned regions of parameter space, but for $\lambda_{\text{eff}} = 1$ can be as high as $S \sim 200$ for the masses plotted in Figs. 1 and 2. In these resonance regions the lower bound on $m_\chi$ can be increased by over an order of magnitude, as shown by the dotted green line in Fig. 1 which shows the bound assuming the strongest possible Sommerfeld enhancement. The effect is reduced for smaller $\lambda_{\text{eff}}$, but for $\lambda_{\text{eff}} = 0.1$ the maximum enhancement is still $S \sim 86$. In regions where this enhancement is realized indirect detection experiments are still sensitive to the model even at couplings $\lambda_{\text{eff}} \lesssim 0.1$, as shown in Fig. 2.

## V. DISCUSSION

Null results at the LHC may indicate that we live in a universe with some degree of fine-tuning, and a mechanism that solves the hierarchy problem may not appear until a scale of order 10–100 TeV. If this is the case, it is a natural expectation that dark matter may have a mass that falls in this range with $\mathcal{O}(1)$ couplings to the SM. DM in this mass range is testable at direct detection experiments and from gamma ray production in the galactic center. However, the usual mechanisms for DM production in the early universe make it difficult to generate the observed dark matter abundance for this mass range and coupling strength. In this work we consider the nonthermal production of dark matter and show that this represents a viable production mechanism for dark matter in this region of parameter space, but requires DM to be nonrelativistic at the reheat temperature. In a conventional cosmology this requires some assumptions about the reheat temperature and potentially creates difficulties accommodating a baryogenesis mechanism. The avoided deconfinement model circumvents these issues, naturally providing dark matter candidates in this mass range that are produced nonthermally with $\mathcal{O}(1)$ couplings to the SM without any strong requirements on the reheat temperature.

---

[3]By maximum possible enhancement we mean the largest enhancement factor for the given coupling in the mass range displayed on the figures.

## ACKNOWLEDGMENTS

I thank Prateek Agrawal for comments on a draft of this manuscript. I also thank Prateek Agrawal, Subir Sarkar, and John March-Russell for useful discussions. M. N. is funded by a joint Clarendon and Sloane-Robinson scholarship from Oxford University and Keble College.

## APPENDIX A: FREEZE-IN CALCULATION

To extract the parametric dependence, one can use the asymptotic form of the Bessel function

$$K_1(z) \sim \sqrt{\frac{\pi}{2z}} e^{-z}, \tag{A1}$$

and change variables to $y = \sqrt{s}/2m_\chi - 1$, giving

$$n_{\text{eq}}^2 \langle \sigma v \rangle = \frac{g_\chi^2 \lambda^2 T^4 x^{1/2}}{128\pi^{9/2}} \int_0^\infty dy\, e^{-2x(1+y)} \sqrt{y(y+1)(y+2)}. \tag{A2}$$

Given the exponential factor in the integrand the integral is dominated for small $y$, so the replacement

$$\sqrt{y(y+1)(y+2)} \to \sqrt{2y} \tag{A3}$$

can be made, leading to the result

$$n_{\text{eq}}^2 \langle \sigma v \rangle \simeq \frac{g_\chi^2 T^4}{256\pi^4} \lambda^2 x e^{-2x}. \tag{A4}$$

The difference between the approximate solution (A4) and the full integral done numerically is found to be less than 5% for $x \gtrsim 5$, with the approximation improving for larger $x$.

## APPENDIX B: THERMAL EQUILIBRIUM

The results of Secs. II and III rely on the fact that $\chi$ never thermalizes. Here we present a cross-check of this assumption. The interactions of $\chi$ with the thermal bath will be too slow to maintain equilibrium if

$$\frac{n_{\text{eq}} \langle \sigma v \rangle}{H} < 1. \tag{B1}$$

For the nonthermal production mechanism considered in Sec. II this condition can be written as

$$\frac{m_\chi}{M_P} > 3.7 \times 10^{-5} \times \lambda^2 e^{-x_r} x_r^{1/2}, \tag{B2}$$

after setting $g_* = 106.75$. Using Eq. (14) to fix $\Omega_\chi h^2 = 0.1$ implies $\lambda^2 e^{-2x_r}(1 + 2x_r) = 4.13 \times 10^{-20}$, and Eq. (B2) provides the strongest bounds for large $\lambda$. Setting $\lambda = 1$ this condition implies $m_\chi \gtrsim 5$ TeV with the bound relaxed for smaller coupling.

For the AD model considered in Sec. II it is more natural to recast Eq. (B2) in terms of the reheat temperature

$$\frac{T_r}{M_P} > 3.7 \times 10^{-5} \times \frac{\lambda^2 e^{-x_r}}{x_r^{1/2}}, \tag{B3}$$

again setting $g_* = 106.75$. The bounds in this case are weaker due to the possibility of a large reheat temperature with $x$ remaining approximately constant at large temperatures, as well as the $1/\epsilon$ enhancement of the final abundance. Setting $\lambda = 1$ and using Eq. (22) to fix $\Omega_\chi h^2 = 0.1$, the bounds on $T_r$ are rather mild,

$$T_r \gtrsim 10 \text{ GeV}, \tag{B4}$$

and get weaker for smaller $\lambda$.